\documentstyle[12pt]{article}
\begin{document}
%\begin{document}

{\bf INTERNAL-CYCLE VARIATION OF SOLAR DIFFERENTIAL ROTATION}  \\

{\it K. J. LI$^{1,2}$, J. L.  XIE$^{1,3}$,  X. J. SHI$^{1,3}$} \\
$^{1}$National Astronomical Observatories/Yunnan Observatory,
      CAS, Kunming 650011, China \\
$^{2}$Key Laboratory of Solar Activity, National Astronomical
Observatories, CAS, Beijing 100012, China  \\
$^{3}$Graduate School of CAS, Beijing 100863, China  \\

{\bf Abstrct.} The latitudinal distributions of the yearly mean rotation rates measured respectively by Suzuki in 1998 and 2012 and Pulkkinen $\&$ Tuominen in 1998 are utilized to investigate internal-cycle variation of solar differential rotation. The rotation rate at the solar Equator seems to decrease since cycle 10 onwards. The coefficient $B$ of  solar differential rotation, which represents the latitudinal gradient
of rotation, is found smaller in the several years after the minimum of a solar cycle than in the several years after the maximum time of the cycle, and  it peaks several years after the maximum time of the solar cycle.  The internal-cycle variation of the solar rotation rates looks similar in profile to that of the coefficient $B$. A new explanation is proposed to address such a solar-cycle related variation of the solar rotation rates.
Weak magnetic fields may more effectively reflect differentiation at low latitudes with high rotation rates than at high latitudes with low rotation rates, and strong magnetic fields may more effectively repress differentiation at relatively low latitudes than at high latitudes.
The internal-cycle variation is inferred to the result of both the latitudinal migration of the surface torsional pattern and the repression of strong magnetic activity to differentiation.  \\
{\bf Sun: rotation-- Sun: activity-- Sun: sunspots}

\section{INTRODUCTION}
The Sun's atmosphere is found to rotate
faster at the equatorial region than at higher latitude regions. In a specific word, it rotates in a circular course by 26 days at the solar Equator but 30 days at $60^{\circ}$ latitude, which is the so-called
differential rotation (Balthasar $\&$ W$\ddot{o}$hl 1980; Gilman $\&$ Howard 1984; Sheeley,  Wang $\&$  Nash 1992; Rybak 1994;  Altrock 2003; Song $\&$ Wang 2005; Chu et al. 2010;  W$\ddot{o}$hl et al. 2010; Li et al. 2013).
Two main methods have been usually exploited to measure rotation velocity of the solar atmosphere: the tracer method and the spectroscopic

method (Howard, Gilman $\&$  Gilman 1984; Pulkkinen $\&$ Tuominen 1998a;
Braj$\breve{s}$a et al.  2000, 2002;   W$\ddot{o}$hl $\&$ Schmidt 2000; Le Mouel et al. 2007; Li et al. 2012). In the solar interior, solar rotation
rate is also determined but by a specialized method:
the helioseismology measurement method (Howe et al. 2000a, 2000b; Antia $\&$  Basu 2001), and the latitudinal migration of rotation angular velocity in solar interior  (Howe et al. 2009) is  found similar to the
torsional oscillation pattern of the solar atmosphere
measured by the spectroscopic method (Howard $\&$ LaBonte 1980; LaBonte $\&$ Howard 1982; Schr$\ddot{o}$ter 1985). At the present, observations and studies for the solar differential rotation have taken  a great achievement (Howard 1984; Schr$\ddot{o}$ter 1985; Snodgrass 1992;  Paterno 2010;  Li et al. 2012)
 However, there are still some aspects, for example, the solar-cycle related and long-term variations of solar rotation rate, unknown now (Komm, Howard $\&$ Harvey 1993; Ulrich $\&$ Bertello 1996;  Stix 2002;
 Li et al., 2011a, 2011b).

In this study, we will investigate internal-cycle variation of solar differential rotation, using
 the measurements of rotation rates made by Suzuki (1998, 2012) and Pulkkinen $\&$ Tuominen (1998a).
 A new explanation is proposed to address such a solar-cycle related variation of the solar rotation rates.

\section{INTERNAL-CYCLE VARIATION OF SOLAR DIFFERENTIAL ROTATION}
\subsection{Revisiting the Measurements of Solar Differential Rotation taken by Suzuki (1998, 2012)}
Daily photographic observations of sunspots at the solar full disk have been made with a refractor of
102 mm aperture and 1200 mm focal length by Suzuki since the year of 1988, and the positions of sunspot groups at the solar disk are obtained (Suzuki 1998, 2012). He measured annual mean rotation rates of sunspots in $5^{0}$ latitude bin during cycles 22 to 23 (from 1998 to 2006) through analyzing his observational data of sunspots. The measurements are given in the Table 1 of Suzuki (1998) for the years of 1988 to 1995 and in the Table 1 of Suzuki (2006) for the years of 1996 to 2006.

The solar differential rotation is usually expressed by the
standard formula (Newton $\&$ Nunn 1951):
$$
\omega (\phi)=A+B sin^{2}\phi
$$
where $\omega (\phi)$ is the solar
sidereal angular velocity at latitude $\phi$ and the coefficients $A$ and
$B$ represent the equatorial rotation rate and the latitudinal gradient
of the rotation, respectively (Howard 1984).
The latitudinal distribution of annual mean  sidereal rotation rates measured by Suzuki (1998, 2012) is fitted by the formula
for each year. Figure 1 shows the cross-correlation coefficient of the formula fitting to the latitudinal distribution of the annual mean rotation rates, and the corresponding tabulated  value at the $95\%$ confidence level is also given. The calculated correlation coefficient is larger than the corresponding tabulated  value for all years except the years of 1993 and 1994, indicating that the formula can statistically significantly give a good fitting to the Suzuki's measurements of annual mean rotation rates. In these two years (1993 and 1994), the correlation coefficients are both less than the corresponding tabulated  values at the $95\%$ confidence level.
Thus for these two years, the fitting values of $A$ and $B$ are replaced by the linearly extrapolated values of their neighboring two points. Figure 2 shows the obtained coefficients $A$ and $B$. A linear fitting is taken to the coefficient $A$ varying with time $t$ (in years),  and  resultantly, $A=39.0713-0.0122\times t$,  and the correlation coefficient is 0.4246, which is statistically significant at the $92\%$ confidence level. There is a decrease trend for $A$, and the decrease rate is $0.0122^{0}/day$ per year. A 5-point smoothing is taken to the coefficient $B$, which is shown in the figure. A special feature for $B$ is that its absolute value is larger in the several years after the minimum of a solar cycle than in the several years after the maximum time of the cycle, and the absolute value of $B$ clearly decreases and then increases within the falling part of a sunspot cycle.
As the figure shows,  a short-term effect can be seen for $B$ varying within a solar cycle, and a
long-term effect, for $A$ trending to decrease.
 The correlation coefficient of yearly sunspot numbers with  $B$ is calculated to be 0.521, which is statistically significant at the confidential level of $95\%$.
While the correlation coefficient of yearly sunspot numbers with  $A$ is  -0.384, which is of no significance.
For the solar surface rotation rate at the solar Equator (the coefficient $A$),
there exists a secular decrease of statistical significance even since Cycle 12 onwards (Javaraiah,  Bertello $\&$ Ulrich 2005a, 2005b; Li et al. 2013).

Based on the obtained coefficients $A$ and $B$, we calculate latitudinal distribution of rotation rates in each year of 1988 to 2006 through the standard formula, which is shown in Figure 3. Rotation rates  are found to obviously show a migration within a solar cycle, but such a migration is different at the falling part of a solar cycle from both the latitudinal migration of sunspots and the torsional oscillation pattern of solar surface differential rotation (Snodgrass 1987; Li et al. 2008).
Figure 4 show four isopleth lines of  rotation rates and their corresponding 5-point smoothing lines.
The isopleth values of rotation rates are 14.4, 14.2, 14.0, and $13.8^{o}\ day^{-1}$  in turn from low to high latitudes. Shown also in the figure are the minimum and maximum times of sunspot cycles.
As Figures 3 and 4 show, the migration of rotation rates seems thwarted at the falling phase of a solar cycle,
compared with the oscillation pattern of solar surface differential rotation (Snodgrass 1987; Li et al. 2008).
It is inferred that strong magnetic fields should repress solar differential rotation, in agreement with Braj$\breve{s}$a et al.  (2006) and  W$\ddot{o}$hl  et al. (2010).

\subsection{Revisiting the Measurements of solar differential rotation taken by Pulkkinen $\&$ Tuominen (1998a)}
The Royal Observatory in Greenwich  started sunspot observations, the long-lasting Greenwich
Photoheliographic Results (GPR) in 1874, lasting about 103 years, and it stopped in 1976. After that, one
of the still continuing records of sunspots has been made by
the Solar Optical Observing Network (SOON) of the US Air
Force together with the US National Oceanic and Atmospheric
Administration (NOAA) (Pulkkinen $\&$ Tuominen 1998a). These GPR and SOON/NOAA data
sets can  be found at the internet web site\footnote{http://wwwssl.msfc.nasa.gov/ssl/pad/solar/greenwich.htm}.
 Pulkkinen $\&$ Tuominen(1998a) used these GPR and SOON/NOAA data  in cycles 10 to 22 namely at the time interval of 1874 to 1996 to study velocity structures from sunspot statistics. They gave in their Figure 3 the rotation profiles of the  data at the equatorial range between latitudes $\pm 20$ degrees at different phases of cycle.
 These rotational velocity values are the sidereal ones.
 The figure is 3-times enlarged, then all points of the rotation profiles  in the figure are measured independently by each of the all three authors of this study, and finally, the measurements given by the three authors are averaged.
Three times of measurements ensure the obtained data matching the original well.
The obtained rotation profiles are fitted here by the standard formula of differential rotations
for each year within a solar cycle, corresponding to different phases of a Schwabe cycle.
A profile gives a set of fitting parameters, reducing the effect of measurement errors of individual points on the fitting parameters.
Figure 5 shows the cross-correlation coefficients of the formula fitting to the rotation profiles and the corresponding tabulated  values at the $98.5\%$ confidence level.
The calculated correlation coefficients are larger than the corresponding tabulated  values for all years,  indicating that the formula can statistically significantly give a good fitting to the annual rotation profile.
The correlation coefficient is lowest in and
around solar activity minima due to the lower number of sunspots
on the Sun in those time intervals.
Figure 6 shows the obtained coefficients $A$ and $B$  at different phases of a Schwabe cycle. A linear fitting is taken to the coefficient $A$ varying with time $t$ of a solar cycle,  and  resultantly, $A=14.6261-0.0107\times t$,  and the correlation coefficient is 0.6029, which is statistically significant at the $93\%$ confidence level. There is a decrease trend for the coefficient $A$, and the decrease rate is $0.0107^{0}/day$ per year. A 3-point smoothing is taken to the coefficient $B$, which is shown in the figure. The same special feature for $B$ is obtained as that obtained through analyzing the data of Suzuki (1998, 2012).

Based on the fitting values of the coefficients $A$ and $B$, we also calculate latitudinal distribution of rotation rates in each year of a solar cycle through the standard formula, which is shown in Figure 7. Rotation rates  are found to obviously show the same migration within a solar cycle as mentioned above.
Figure 8 shows four isopleth lines of  rotation rates and their corresponding 3-point smoothing lines.
The isopleth values of rotation rates are 14.2, 13.9, 13.6, and $13.3^{o}\ day^{-1}$  in turn from low to high latitudes. As Figures 7 and 8 show, large sunspots seem to hinder the torsional oscillation pattern  shifting towards the Equator  $1\sim 2$ years later after the start of a solar cycle (Snodgrass 1987; Li et al. 2008).
The strong magnetic fields should obviously repress differentiation of rotation rates.

\section{CONCLUSIONS AND DISCUSSION}
The latitudinal distributions of the yearly mean rotation rates measured respectively by Suzuki (1998, 2012) and Pulkkinen $\&$ Tuominen (1998a) are exploited to investigate internal-cycle variation of solar differential rotation. Firstly , they are fitted by the standard formula of solar differential rotation. Resultantly,
the rotation rate at the solar Equator is then found to decrease since cycle 10 onwards, and
the decrease rate is  about $0.011^{0}/day$ per year within a solar cycle. For the coefficient $B$,
its absolute value is found larger in the several years after the minimum of a sunspot cycle than in the several years after the maximum time of the cycle, and the absolute value  clearly decreases and then increases within the falling part of the sunspot cycle, namely, the coefficient $B$   peaks several years after the maximum time of the solar cycle. Such a profile of the coefficient $B$ in a solar cycle was also given in the Figure 5 of Javaraiah (2003). Although the variations of the coefficients $A$ and $B$  within a solar cycle obtained through analyzing the data of   Pulkkinen $\&$ Tuominen (1998a) are similar to those through analyzing the data of Suzuki (1998, 2012), the former should be more plausible, because the data utilized by Pulkkinen $\&$ Tuominen (1998a) are more reliable in observations and much longer in time than by Suzuki (1998, 2012). Thus at follows we mainly focus attention on the former.

The differential rotation of solar atmosphere has a periodical  pattern of
change. Such a pattern can be described by the so-called
torsional oscillation, in which the solar differential rotation should be cyclically
speeded up or slowed down in certain zones of latitude
while elsewhere the rotation remains essentially steady (Snodgrass $\&$ Howard 1985; Li et al. 2008; Li et al. 2012). The zones of anomalous rotation move on the Sun in wavelike
fashion, keeping pace with and flanking the zones of
magnetic activity (LaBonte $\&$ Howard, 1982; Snodgrass $\&$ Howard, 1985).
The surface torsional pattern, and perhaps the magnetic activity
as well, are only the shadows of another unknown phenomenon
occurring within the convection zone (Snodgrass 1987; Li et al. 2008).
In this investigation, the internal-cycle  variation of solar differential rotation
can be explained as follows on the framework of the surface torsional pattern, which is briefly stated above,
and by the magnetic activity together.

Braj$\breve{s}$a,  Ru$\breve{z}$djak $\&$ W$\ddot{o}$hl (2006) once investigated  solar-cycle related variations of solar rotation rate. They found a higher than average rotation velocity in the minimum time of a Schwabe cycle, and  a plausible interpretation was then given. When magnetic fields are weaker, one can expect a more pronounced
differential rotation yielding a higher rotation velocity at low latitudes on an average (Braj$\breve{s}$a,  Ru$\breve{z}$djak $\&$ W$\ddot{o}$hl 2006). As Figures 2 and 6 shows, more pronounced differentiation of rotation rates appears indeed at the first 4 years of a solar cycle than at at the second 4 years, due to that weaker magnetic fields should appear at the first 4 years. Strong magnetic fields should repress differentiation, but weak magnetic fields seem to just reflect differentiation of rotation rates. Further, weak magnetic fields may more effectively reflect differentiation at low latitudes with high rotation rates than at high latitudes with low rotation rates, and strong magnetic fields may more effectively repress differentiation at relatively low latitudes than at high latitudes.
As Figures 2 and 6 display, the coefficient $B$ may be divided into three parts. Part one spans from the start to the $4^{th}$ year of a solar cycle, the absolute $B$ is approximately a constant or slightly fluctuates.
Relatively high latitudes and relatively weak magnetic fields at this time interval make the repression action of sunspots less obvious than after this interval.
Part two spans from
the $4^{th}$  to the $7^{th}$ year, the absolute $B$ decreases. When solar activity is progressing into this part, sunspots appear at lower and lower latitudes, magnetic fields repress differentiation more and more effectively, and differentiation appears less and less conspicuously, thus the absolute $B$ decreases within this part. Part three spans from the $7^{th}$ year to the end of a solar cycle.  Within this part, magnetic fields become more and more weak, they repress differentiation  less and less effectively, and sunspots  appearing at more and more low latitudes lead to that the differentiation reflected by latitudinal migration should be more and more conspicuous, thus, the absolute $B$ increases (Li et al. 2012). In sum, the internal-cycle variation of solar differential rotation
is inferred to the result of both the latitudinal migration of the surface torsional pattern and the repression of strong magnetic activity. It means that measurements of differential rotation should different at different phase of a Schwabe cycle or/and at different latitudes (different spacial positions of observed objects on the solar disk), that is the main reason why too many different results about solar differential rotation exist  at the present
(Howard 1984; Schr$\ddot{o}$ter 1985; Snodgrass 1992; Beck 1999;
Paterno 2010; Li et al. 2012).

The solar differential rotation is not a fossil but is proposed generated and continuously
maintained by the angular momentum transport from higher
latitudes toward the Equator (Pulkkinen $\&$ Tuominen 1998a). Measurements from the GPR by
Ward (1965) revealed the existence of this transport, mostly by the
Reynolds stress. Vr$\breve{s}$nak et al. (2003) found a statistically significant correlation
between rotation residuals and meridional motions, through tracing ``point-like structures'' (predominantly young coronal bright points), indicating that the existence of the equatorward transport
of angular momentum. Tracing of coronal bright points (CBPs) provides an extension of
the Reynolds stress analysis to high latitudes which could
be an important tool to investigate the dependence of the
Reynolds stress on the latitude (Vr$\breve{s}$nak et al. 2003), because CBPs are interesting features to be used for the rotation estimation
along a solar cycle since they appear even on the full disk and at both minimum and
maximum cycle phases (Zaatri et al. 2009; W$\ddot{o}$hl et al. 2010).
The torsional oscillation that was found by
Howard and LaBonte (1980) indicated that latitudinal motions should exist, and they may result both
from hydrodynamic circulation and solar magnetic cycle, therefore the Reynolds stress is strongly
present (Tuominen, Tuominen $\&$ Kyr$\ddot{o}$l$\ddot{a}$inen 1983; R$\ddot{u}$diger et al. 1986; Pulkkinen et al. 1993; Pulkkinen $\&$ Tuominen 1998a). The horizontal Reynolds stress should be a function (dependence) of not only the gradients of rotation but also rotation itself (Pulkkinen $\&$ Tuominen 1998a).
We propose here that the spacial variations of the solar magnetic activity act on differential rotations in different ways (see the explanation on Figures 2 and 6). It means that the latitudinal shift velocity should be related with
differential rotations, accordingly supporting this dependence.

Comparison of Figure 8 with Figure 6 shows that the internal-cycle variation (isopleth lines) of the solar rotation rates looks similar in profile to that of the coefficient $B$, and the former is mainly reflected by the latter.
The strong magnetic fields should hinder the torsional oscillation pattern  migrating towards the Equator, and  a block is thus formed in the migration ``river".  The block peaks when the $B$ peaks. The internal-cycle variation of the solar rotation rates is inferred to the result of both the latitudinal migration of the surface torsional pattern and the repression of strong magnetic activity to differentiation. Indeed the surface torsional pattern, and perhaps the magnetic activity as well, are only the shadows of another unknown phenomenon
occurring within the convection zone (Snodgrass 1987; Li et al. 2008).

Based on the above explanation about the variation profile of $B$ within a solar cycle, the temporal distribution of
magnetic activity strength in a solar cycle should give rise to $B$ periodically distributed in a solar cycle, but the spacial distribution  of magnetic activity strength should disturb $B$ to form such a solar-cycle-period distribution. Therefore, the variation profile of the coefficient $B$ in a solar cycle (see Figures 2 and 6) is
 similar to but obviously different from the variation profile of sunspot numbers in a solar cycle. This seemingly implies that the coefficient $B$ should possibly have a not-too-strong or weak relation with sunspot activity at the scale of solar cycles. We calculated the correlation coefficient of yearly sunspot numbers respectively with the coefficients $A$ and $B$ determined through the data of Suzuki (1998, 2012), and resultantly it  is 0.521 for $B$, which is statistically significant at the confidential level of $95\%$. But for $A$ it is -0.384, which is of no significance. However, Jurdana-$\breve{S}$epi$\acute{c}$ et al. (2011) recently found that the coefficient $A$ is significantly related with sunspot activity, while $B$ is not related, through tracing small bright coronal structures.
   Chandra et al. (2010) found that $B$ should not show any systematic variations  for the soft X-ray corona, which is in agreement with the resutls of  Jurdana-$\breve{S}$epi$\acute{c}$ et al. (2011).
   The correlation of sunspot activity  with the rotation coefficients determined by tracing sunspots contradicts that determined by tracing coronal activity events. The possible reason is inferred to be that the coronal magnetic field is much weaker than the sunspot magnetic field. For the time scale longer than solar cycles, the coefficient $A$ is found negatively correlated with sunspot activity, while $B$ should be hardly correlated (Javaraiah,  Bertello $\&$ Ulrich 2005a, 2005b; Javaraiah $\&$ Ulrich 2006; Chandra $\&$ Vats 2011; Li et al. 2012).

\vskip 1in

We thank the anonymous referees for their careful reading of the
manuscript and constructive comments which improved the original
version of the manuscript. This work is supported by the 973
programs 2012CB957801 and 2011CB811406, the
National Natural Science Foundation of China (11273057, 11221063, 11147125, and 11073010),  and the Chinese Academy of Sciences.

\clearpage

\newpage
\input{epsf}
\begin{figure*}
\begin{center}
\epsfysize=12.cm\epsfxsize=12.cm \hskip -5.0 cm \vskip 30.0 mm
\epsffile{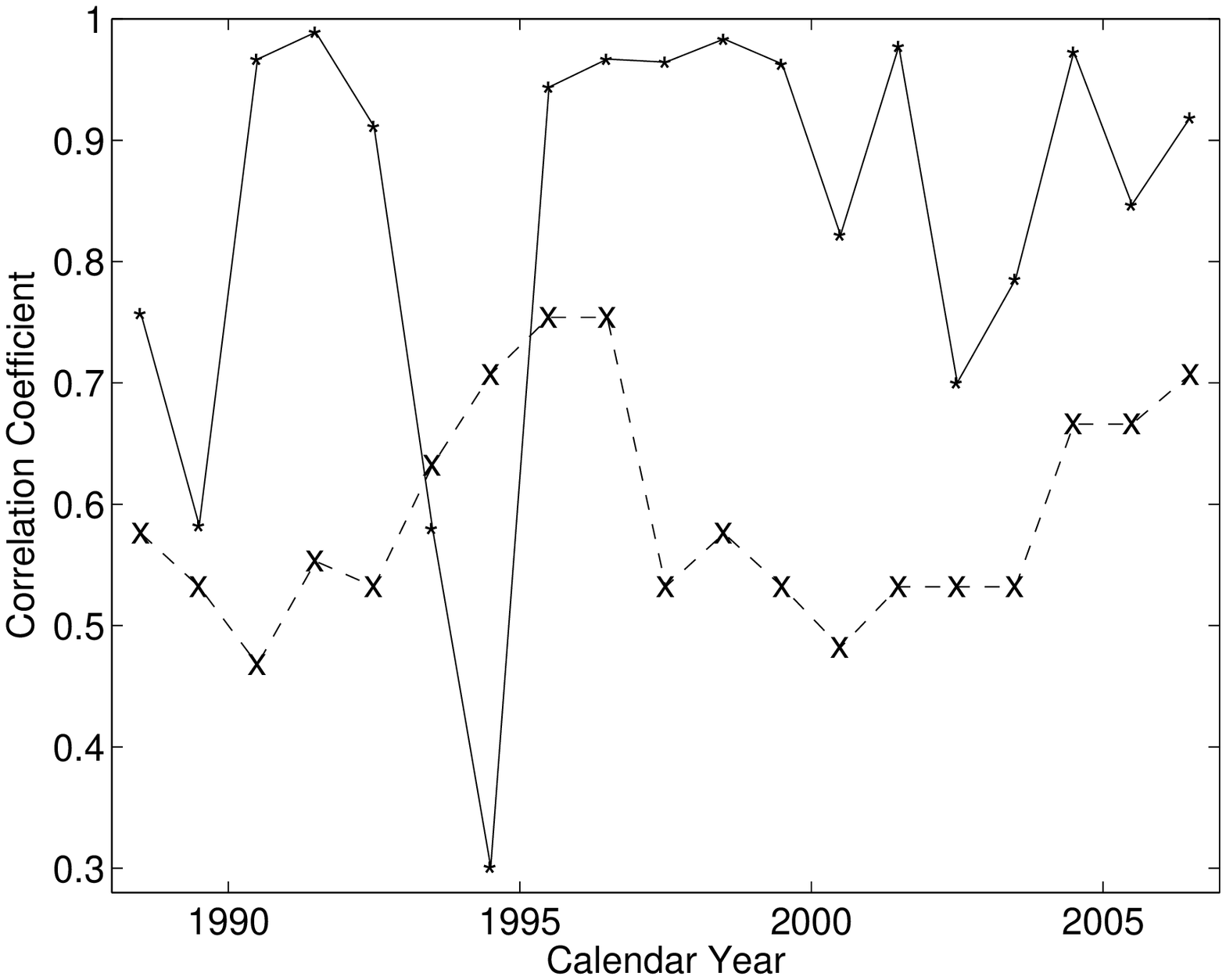} \vskip 2.5cm {{\bf Figure.1}\
The correlation coefficient (the asterisk symbol)
of the latitudinal distribution of annual mean  sidereal rotation rate measured by Suzuki (1998, 2012) at each year of 1988 to 2006 fitted by the standard formula of solar differential rotation.
The cross symbol shows its corresponding tabulated  value at the $95\%$ confidence level.

}
\end{center}
\end{figure*}

\newpage
\input{epsf}
\begin{figure*}
\begin{center}
\epsfysize=12.cm\epsfxsize=12.cm \hskip -5.0 cm \vskip 30.0 mm
\epsffile{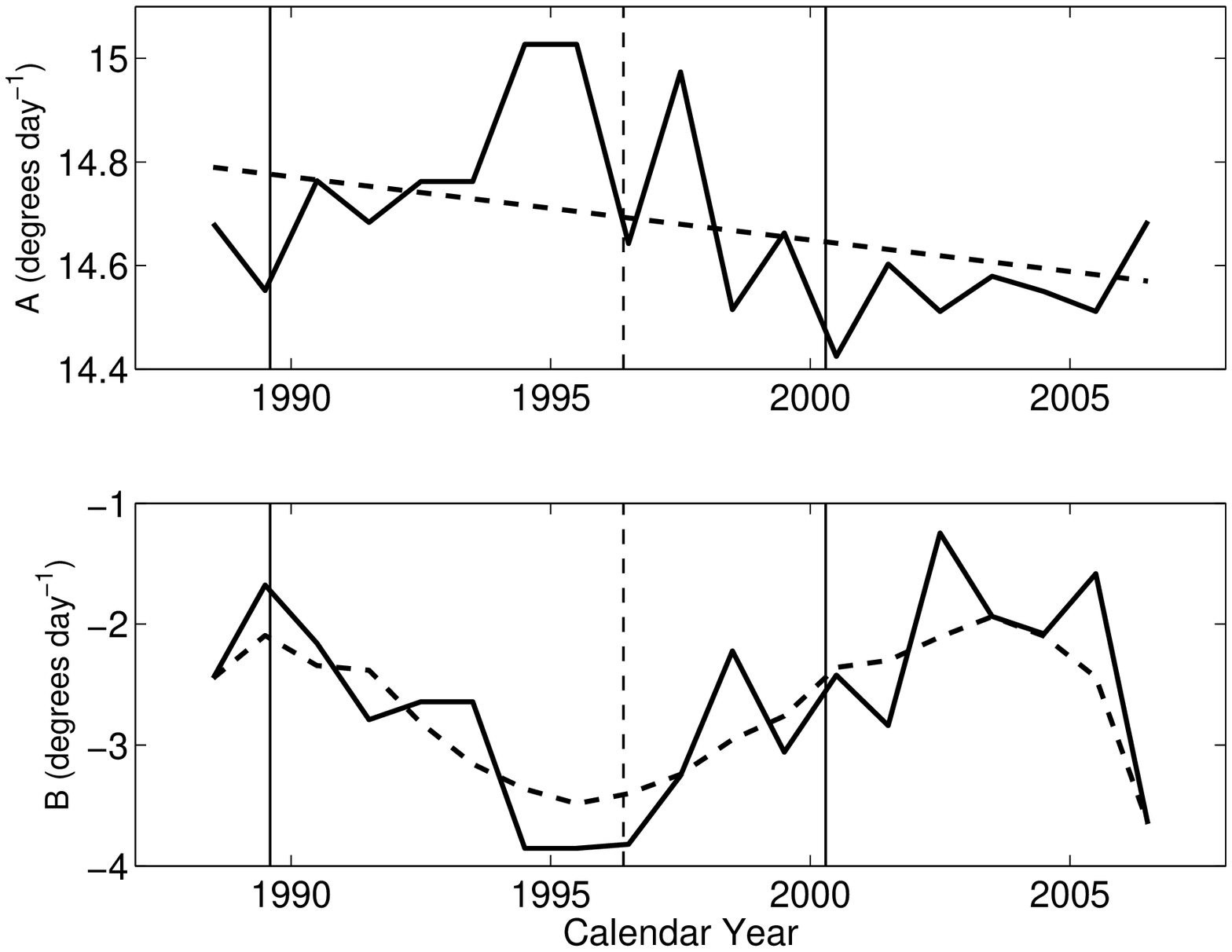} \vskip 2.5cm {{\bf Figure.2}\
The top panel: the coefficient $A$ (the bold solid line)  obtained through fitting the latitudinal distribution of
annual mean sidereal rotation rates  measured by Suzuki (1998, 2012) with the standard formula of solar differential rotation.
The dashed line shows the linear fitting to the obtained coefficient $A$.
The bottom panel: the coefficient $B$ (the bold solid line) obtained through fitting the latitudinal distribution of
annual mean rotation rates  measured by Suzuki (1998, 2012) with the standard formula of solar differential rotation.
The dashed line shows its 5-point smoothing.
In the two panels, the thin vertical dashed/solid lines indicate the minimum/maximum times of sunspot cycles.

}
\end{center}
\end{figure*}

\newpage
\input{epsf}
\begin{figure*}
\begin{center}
\epsfysize=12.cm\epsfxsize=12.cm \hskip -5.0 cm \vskip 30.0 mm
\epsffile{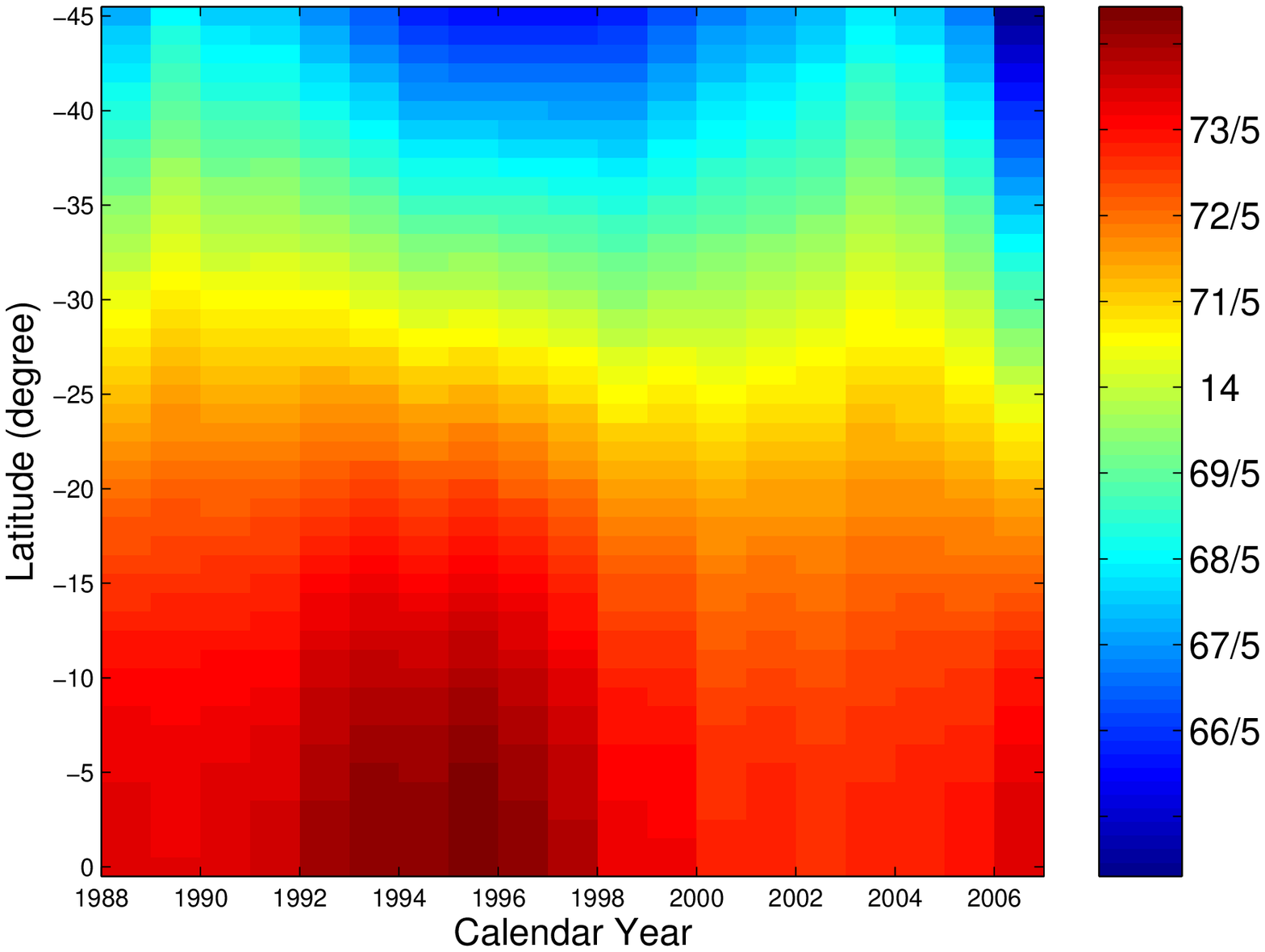} \vskip 2.5cm {{\bf Figure.3}\
Latitudinal distribution of sidereal rotation rates in each year of 1988 to 2006 given through
fitting the latitudinal distribution of
annual mean rotation rates  measured by Suzuki (1998, 2012). The unit shown on the color bar is
$degrees\ day^{-1}$. The fraction numbers at the right side of the figure from the top to the bottom correspond
to the decimal numbers of 14.6, 14.4, 14.2, 14.0, 13.8, 13.6, 13.4, and 13.2 in turn.

}
\end{center}
\end{figure*}

\newpage
\input{epsf}
\begin{figure*}
\begin{center}
\epsfysize=12.cm\epsfxsize=12.cm \hskip -5.0 cm \vskip 30.0 mm
\epsffile{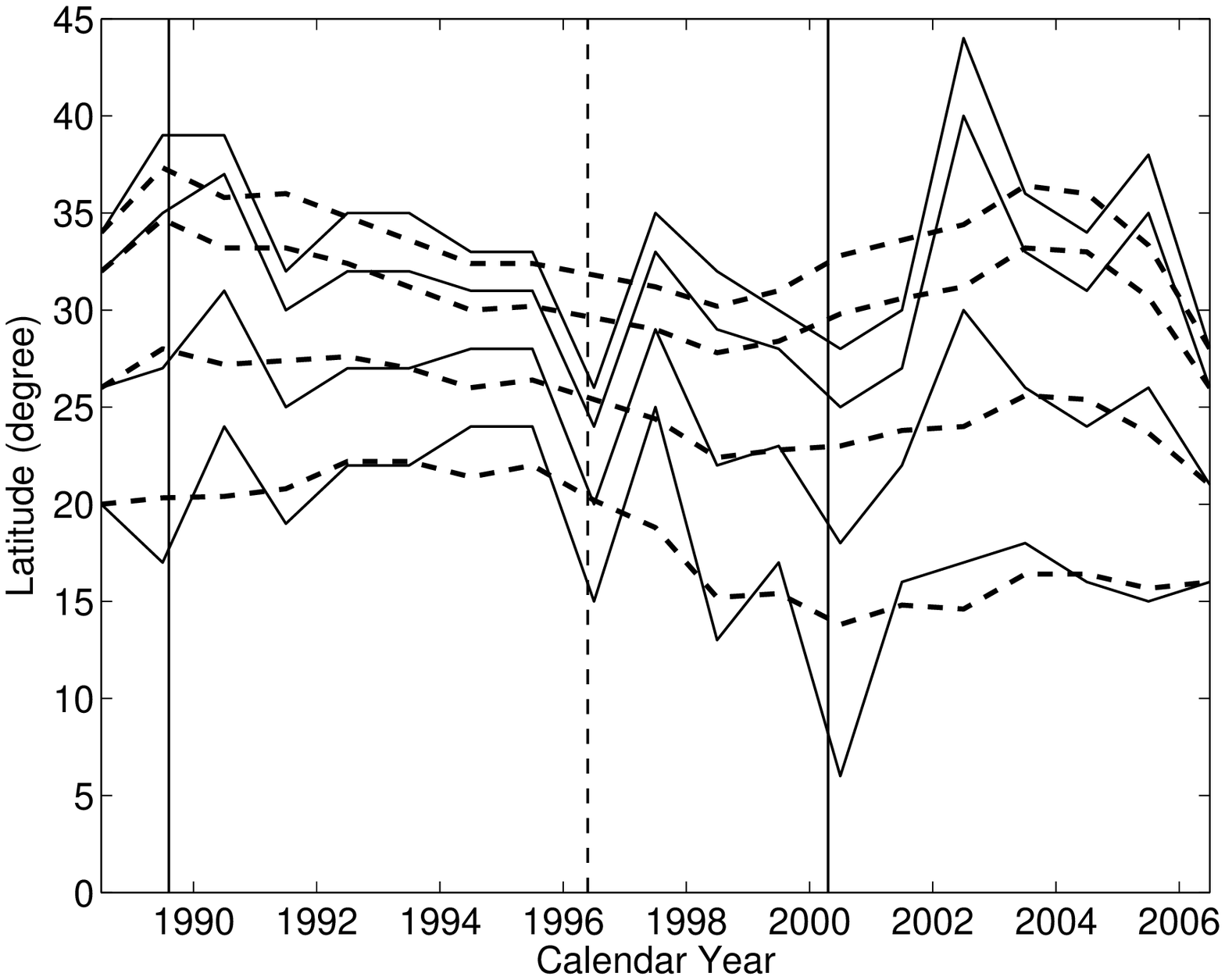} \vskip 2.5cm {{\bf Figure.4}\
Isopleth lines (the solid curves) of the rotation rates shown in Figure 3 and  their corresponding 5-point smoothing lines (the bold dashed curves). The isopleth rotation rates are 14.4, 14.2, 14.0, and $13.8^{o}\ day^{-1}$ in turn form low to high latitudes. The thin vertical dashed/solid lines indicate the minimum/maximum times of sunspot cycles.

}
\end{center}
\end{figure*}

\newpage
\input{epsf}
\begin{figure*}
\begin{center}
\epsfysize=12.cm\epsfxsize=12.cm \hskip -5.0 cm \vskip 30.0 mm
\epsffile{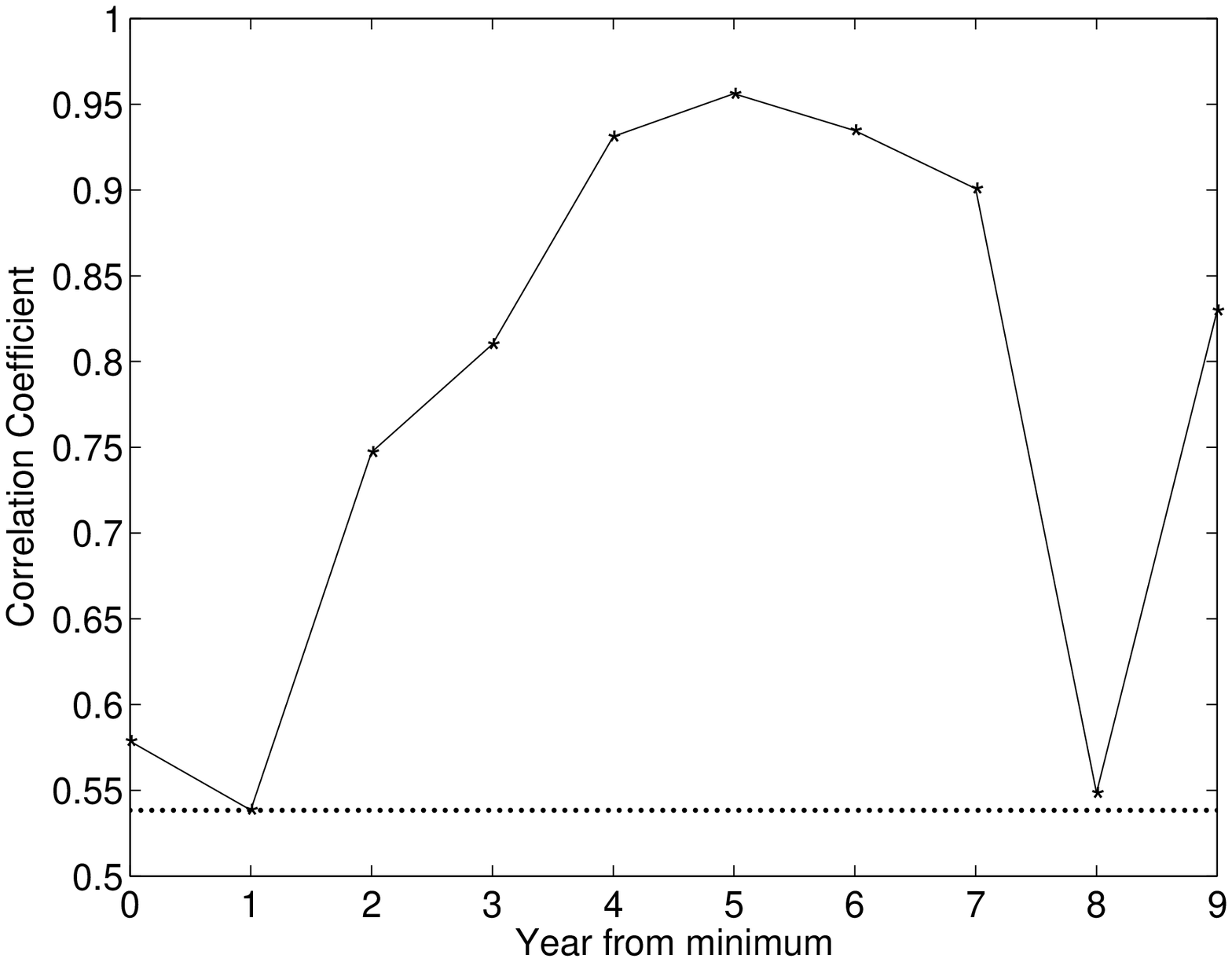} \vskip 2.5cm {{\bf Figure.5}\
The correlation coefficient (the asterisk symbol)
of the latitudinal distribution of annual mean sidereal rotation rate measured by Pulkkinen $\&$ Tuominen (1998a)
at different phases of cycle  fitted by the standard formula of solar differential rotations.
The dotted line shows its corresponding tabulated  value at the $98.5\%$ confidence level.

}
\end{center}
\end{figure*}

\newpage
\input{epsf}
\begin{figure*}
\begin{center}
\epsfysize=12.cm\epsfxsize=12.cm \hskip -5.0 cm \vskip 30.0 mm
\epsffile{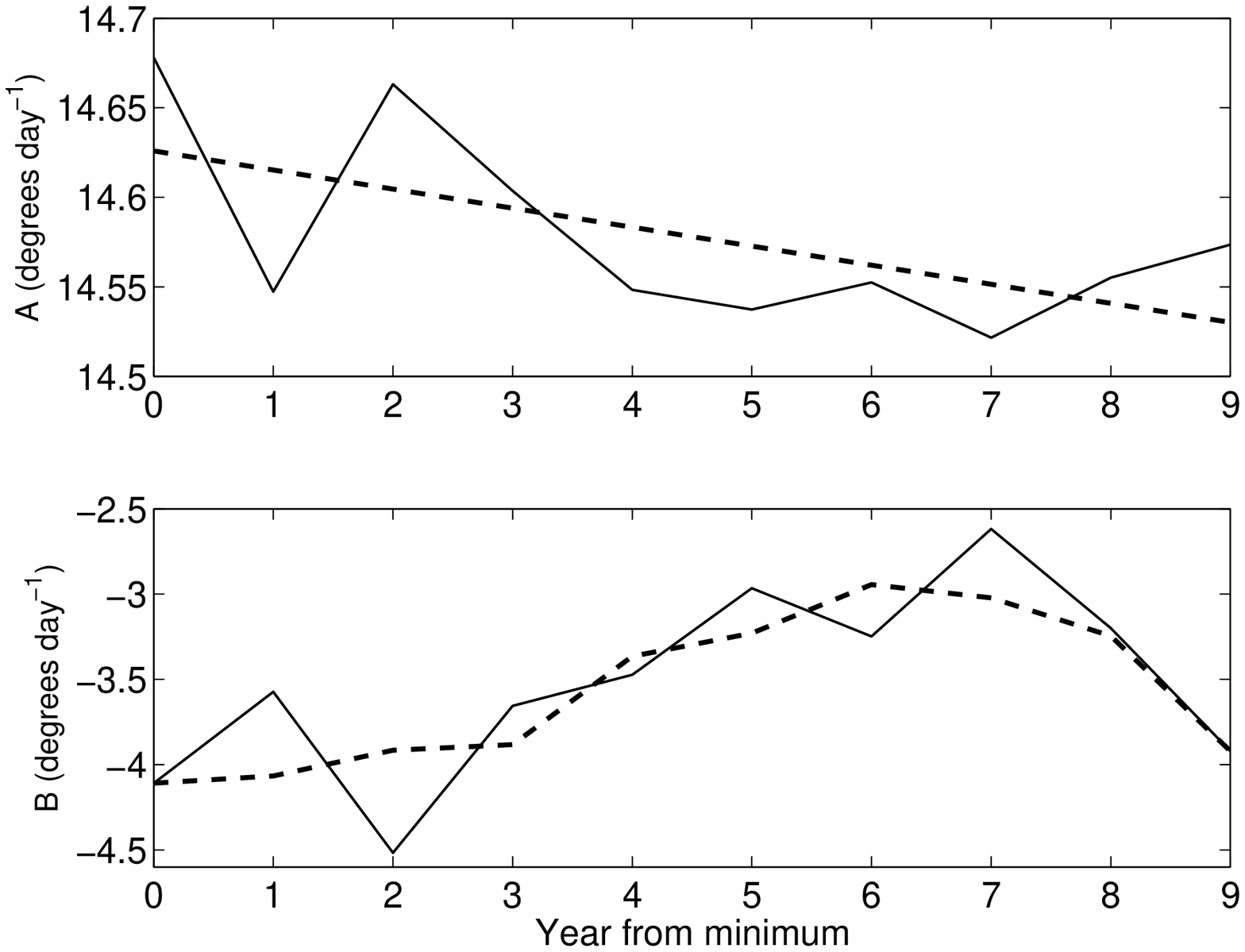} \vskip 2.5cm {{\bf Figure.6}\
The top panel: the coefficient $A$ (the solid line)  at different phases of cycle, which is obtained through fitting the latitudinal distribution of
annual mean sidereal rotation rates  measured by Pulkkinen $\&$ Tuominen (1998a) with the standard formula of solar differential rotations.
The dashed line shows the linear fitting to the obtained coefficient $A$.
The bottom panel: the coefficient $B$ (the  solid line) at different phases of cycle, which is obtained through fitting the latitudinal distribution of
annual mean rotation rates  measured by Pulkkinen $\&$ Tuominen (1998a) with the standard formula of solar differential rotation.
The dashed line shows its 3-point smoothing.

}
\end{center}
\end{figure*}

\newpage
\input{epsf}
\begin{figure*}
\begin{center}
\epsfysize=12.cm\epsfxsize=12.cm \hskip -5.0 cm \vskip 30.0 mm
\epsffile{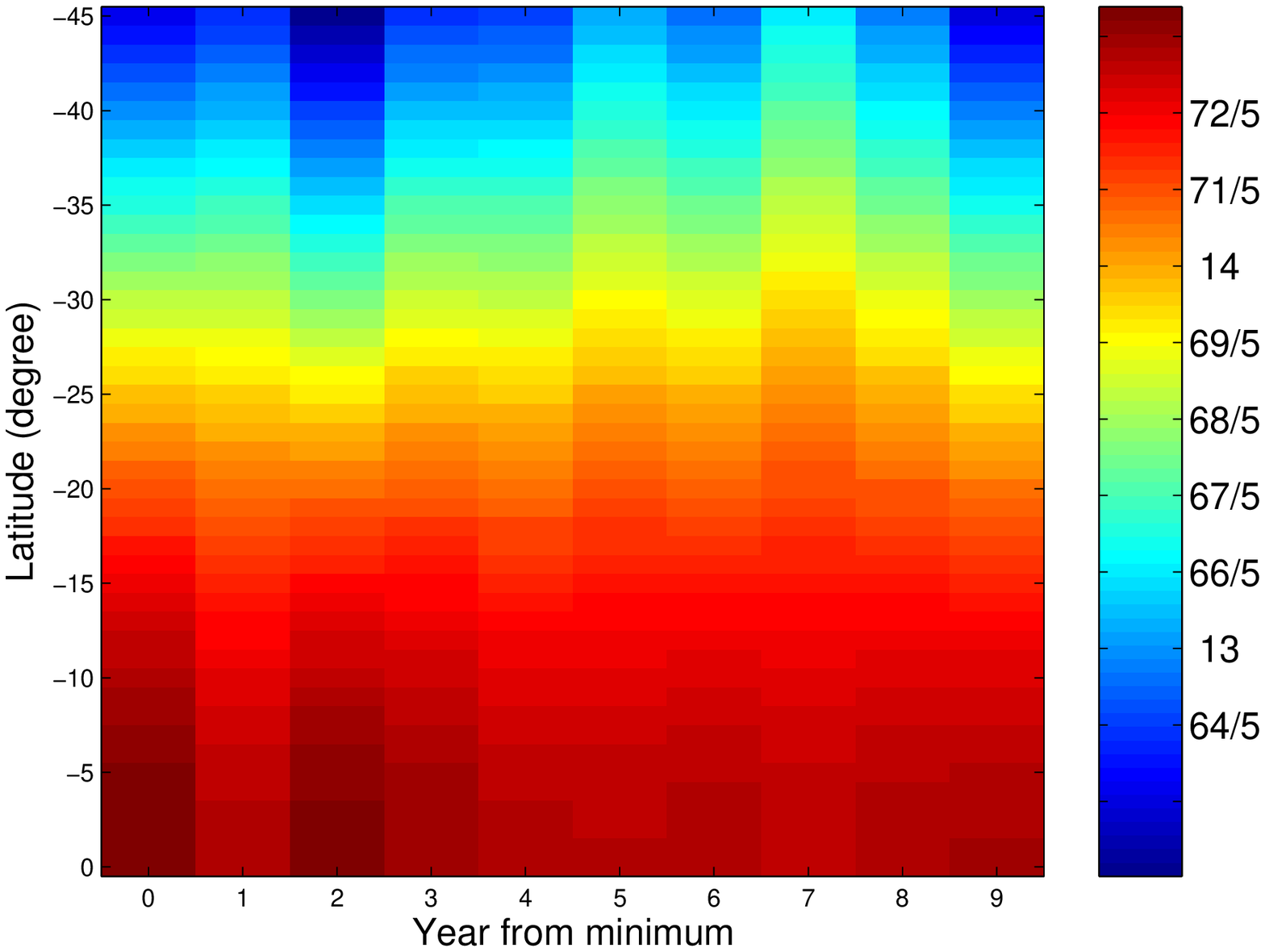} \vskip 2.5cm {{\bf Figure.7}\
Latitudinal distribution of sidereal rotation rates at different phases of cycle, which is given through
fitting the latitudinal distribution of
annual mean rotation rates  measured  by Pulkkinen $\&$ Tuominen (1998a). The unit shown on the color bar is
$degrees\ day^{-1}$. The fraction numbers at the right side of the figure from the top to the bottom correspond
to the decimal numbers of 14.4, 14.2, 14.0, 13.8, 13.6, 13.4, 13.2, 13.0, and 12.8 in turn.

}
\end{center}
\end{figure*}

\newpage
\input{epsf}
\begin{figure*}
\begin{center}
\epsfysize=12.cm\epsfxsize=12.cm \hskip -5.0 cm \vskip 30.0 mm
\epsffile{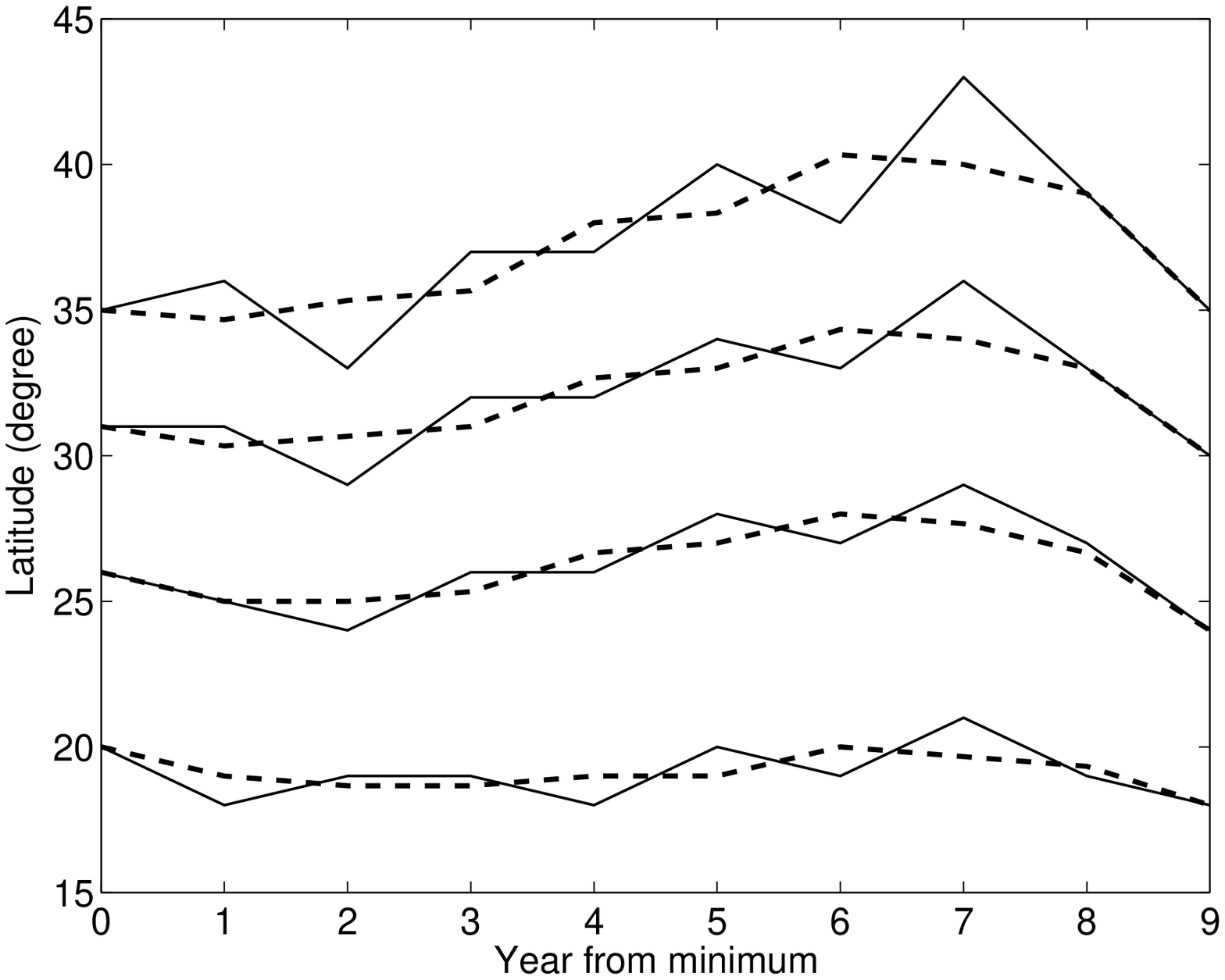} \vskip 2.5cm {{\bf  Figure.8}\
Isopleth lines (the solid curves) of  rotation rates shown in Figure 7 and  their corresponding 3-point smoothing lines (the  dashed lines). The isopleth rotation rates are 14.2, 13.9, 13.6, and $13.3^{o}\ day^{-1}$ in turn form low to high latitudes.

}
\end{center}
\end{figure*}

\end{document}